\documentclass[12pt]{JHEP3}
\usepackage{amsmath,graphicx,cite}
\def\Bbb{\mathbb}
\def\BZ{\Bbb Z} 
\def\BC{\Bbb C} \def\BP{\Bbb P}
 \def\BF{\Bbb F}

\def\BQ{\Bbb Q}
\def\ch{\textrm{ch}}
\def\etal{{\it et~al.}}
\title{Fractional two-branes, toric orbifolds and the quantum McKay correspondence}
\author{Bobby Ezhuthachan\\
Tata Institute of Fundamental Research, Homi Bhabha Rd, Mumbai 400005 India\\
Email: \email{bobby@theory.tifr.res.in}}
\author{Suresh Govindarajan\\
Institut f\"ur Theoretische Physik, ETH, CH-8093 Z\"urich, Switzerland \\
and \\
\mbox{}\thanks{Permanent Address.}
Dept. of Physics, Indian Institute of Technology Madras,
Chennai 600036 India \\ 
Email: \email{suresh@physics.iitm.ac.in}}
\author{T. Jayaraman\thanks{On leave from 
The Institute of Mathematical Sciences,
Chennai. } \\ Dept. of Mathematics,  Tata Inst. of
Fundamental Research, Mumbai India\\
Email: \email{jayaram@imsc.res.in}}
\abstract{We systematically study and obtain the large-volume analogues of 
fractional two-branes on resolutions of the orbifolds $\BC^3/\BZ_n$. We 
also study a 
generalisation of the McKay correspondence proposed in hep-th/0504164 called 
the quantum McKay correspondence by constructing duals to the fractional two-branes. Details are explicitly worked out for two 
examples --
the crepant resolutions of $\BC^3/\BZ_3$ and $\BC^3/\BZ_5$.}
\preprint{{\tt hep-th/0606154}\\
{\tt IITM/PH/TH/2006/1} \\
{\tt TIFR/TH/06-13} \\
June, 2006}
\begin{document}

\section{Introduction}

In an earlier paper \cite{Ezhuthachan:2005jr}, hereinafter
referred to as I, we had initiated the study of fractional
$2p$-branes on non-compact orbifolds. These constructions were a
logical extension of the fractional zero-branes on orbifolds
which have been well studied. While the actual construction of
the boundary states of these fractional $2p$-branes in orbifolds
had been done earlier, there had been no detailed study of these
branes as part of the general study of B-type branes on the
non-compact Calabi-Yau (CY) manifolds in whose K\"ahler moduli
space these orbifolds appeared as special points. In particular
there had been no detailed study of the large-volume analogues of
these fractional $2p$-branes, apart from some useful preliminary
remarks in \cite{Douglas:2000qw}.

While these fractional branes are of interest in their own right,
in I we also showed two interesting connections between these
fractional branes and other constructions in the study of B-type
branes in the Landau-Ginzburg (LG) phase of compact CY manifolds.
First we showed that a particular class of fractional two-branes
in the LG orbifold, on restriction to the CY hypersurface, were
indeed the same as the branes that had been constructed by Ashok
\etal \cite{Ashok:2004zb} using the techniques of boundary
fermions and matrix factorisation of the world-sheet
superpotential. This related the boundary fermion and matrix
factorisation construction to a more straightforward physical
construction in terms of simple boundary conditions on the fields
of the LG model\cite{Govindarajan:1999js}. Second, we argued that
the corresponding conformal field theory boundary states at the
LG point in the K\"ahler moduli space of the CY were in fact the
so-called permutation branes \cite{Recknagel:2002qq} of the
Gepner construction of the bulk world-sheet conformal field
theory. Our argument was further strengthened by the complete
computation presented by \cite{Brunner:2005fv} for these
permutation branes(see also \cite{Enger:2005jk}).

However the construction of the large-volume analogues of the
fractional two-branes in I had relied heavily on some physical
arguments. In particular, the coherent sheaves that corresponded
to these fractional two-branes appeared as the cohomology of
exact sequences that had really had no proper mathematical
meaning, as the assignment of fractional 2-brane charges was
entirely ad hoc. By independent cohomology and K-theory arguments
we had however established that the large-volume analogues of the
fractional two-branes were entirely sensible from the point of
view of sheaves that were associated to the full non-compact CY
manifold. In particular in the full non-compact space the
fractional charge had an entirely sensible interpretation as an
integer charge in a new basis\footnote{The appearance of
fractional charges was already noted in \cite{romel1}}.

In this paper we re-examine these questions from the point of
view of toric geometry. The toric description, as we will see, is
particularly useful in the description of sheaves on the full
non-compact CY. We will show that the charges of the large-volume
analogues of the fractional two-branes, in terms of the Chern
characters of these objects, can be determined consistently in
the framework of toric geometry. We will also see that this
determination shows the heuristic sequences that we wrote down
for these objects are also meaningful in a sense that we will
explain.

Another aspect of these fractional two-branes that we had studied briefly
in I was their relation to the quantum McKay correspondence. One way of
stating the classical McKay correspondence is as a relation between the
sheaves corresponding to the fractional zero-branes and a dual set of
bundles, the so-called tautological bundles, on the full non-compact
CY\cite{orb1,orb2,dd}. For fractional $2p$ branes we conjectured a new
quantum McKay correspondence where a new set of sheaves dual to the
fractional $2p$-branes play the role of the tautological bundles for the
fractional zero-branes. In this paper, we use the toric description to
determine these dual objects carefully for fractional two-branes on
$\BC^3/\BZ_n$ orbifolds.

The organisation of the paper is as follows: In section 2, we
discuss aspects of local Chern characters that one needs to work
with in the context of non-compact manifolds. In sections 3 and
4, we systematically work out the Chern characters for coherent
sheaves associated with fractional zero- and two-branes for the
resolutions of the orbifolds, $\BC^3/\BZ_3$ and $\BC^3/\BZ_5$.
This is done by using the open-string Witten index computed in
conformal field theory as input along with an ansatz for the form
of the Chern character of the coherent sheaves. We fix an
ambiguity that arises due to linear equivalences and then present
concrete objects that reproduce the computed Chern character. In
section 5, we discuss the quantum McKay correspondence and work
out the Chern character for sheaves that are dual to the
fractional two-branes and finally propose candidate objects for
the dual sheaves that are consistent with the computed Chern
character.  We present our conclusions in section 6. Some of the
details of the computations have been presented in four
appendices.

\section{Background}

The gauged linear sigma model has provided a concrete model which
enables one to interpolate between orbifolds and their
resolution. The complexified Fayet-Iliopoulos parameters are the
blow-up moduli and at ``large volume" give the sizes of various
cycles. In the presence of a boundary preserving B-type
supersymmetry, at large volume, the D-branes are best described
as coherent sheaves while at the orbifold point one can construct
boundary states. Relating these two different descriptions has
lead to a surprising connection to the McKay correspondence for
fractional
zero-branes\cite{Govindarajan:2000vi,Tomasiello:2000ym,mayr}.

The main goal of this paper is to obtain a systematic
understanding of coherent sheaves that are obtained by analytic
continuation of fractional two-branes from the orbifold point to
large-volume. The construction of the fractional two-branes as
boundary states in the orbifold is standard. The open-string
Witten index is independent of this analytic continuation and
provides an important input in identifying the relevant coherent
sheaves. However, this data is not enough to reconstruct even the
Chern character (equivalently, the RR charges). An additional
complication is that the fractional two-branes are non-compact
objects and this has to be dealt with as well. As was shown in
paper I, by working in the full non-compact space rather than on
compact sub-manifolds, one obtains an integral basis for the RR
charges carried by the fractional two-branes. In examples with
several divisors having non-trivial intersections, such as the
resolution of $\BC^3/\BZ_5$, the restriction to a particular
compact divisor is not useful either.

As is well known, there is an intimate connection between the
GLSM and toric geometry.  We thus make use of standard
constructions in toric geometry to systematise our study. We also
provide a self-contained description of the necessary details in
the sequel and the appendices. Based on these considerations, we
write a general ansatz for the Chern character of the fractional
two-branes and fix the coefficients using the open-string Witten
index. However, this data is insufficient to fix all
coefficients. We then provide additional input using the
structure of the boundary states that uniquely fixes all
coefficients. Finally, the heuristic method presented in I is
used to provide candidate coherent sheaves that are compatible
with computed local Chern characters.

\subsection{Local Chern characters for sheaves on non-compact CY}

In general, for a variety $X$, the Chern character is a map from the
K-group $K(X)$ to the Chow group with rational coefficients
$A^*(X) \otimes {\BQ}$,
\begin{equation}
\ch: K_0(X) \rightarrow A^*(X) \otimes \BQ\ .
\end{equation}
By standard definitions we may identify
\begin{equation}
A^*(X)\otimes \BQ = \oplus A^p(X) \otimes \BQ = \oplus A_{n-p}
 \otimes \BQ\ .
\end{equation}

Following \cite{Fulton}, it is particularly easy to write down
the Chow group $A_k(X)$ in the case of a toric variety. For a
toric variety $X = X(\Delta)$, where $\Delta$ is the
corresponding fan, $A_k(X)$ is generated by all the classes of
the orbit closures $V(\sigma)$ of $(n-k)$ dimensional cones of
the fan $\Delta$, modulo relations. With the last caveat on
relations, the statement above is true for non-compact as well as
compact varieties.

Now the orbit closures for a given fan $\Delta$ may be simply
written down in terms of the divisors $D_i$, corresponding to the
orbit closure of the one-dimensional cones, as well as the
intersection of these divisors associated to orbit closures of
higher dimensional cones spanned by these one-dimensional cones.
Thus the Chern character involves in general a constant term for
the rank, then terms of the form $D_i$, terms of the form
$D_i\cdot D_j$ and then terms of the form $D_i\cdot D_j\cdot
D_k$. There are no more terms since we are dealing with a complex
threefold. For any given sheaf $E$, these terms have rational
coefficients that have to be determined.

Note that in the case of a non-compact variety, some of these
terms may be in fact zero. The classic example is the case of the
the complex affine spaces, which have only one cohomology, so in
fact the relations in the Chow group set all but one of the $A_k$
to zero. However in our case without explicitly trying to
determine such relations, we will use the fact that in the
intersection form only such triple intersections will survive as
are allowed by the toric construction. This we suppose will
self-consistently determine which terms in the Chern character
will be zero or non-zero. It is easy to see that similar
computation for the affine spaces gives self-consistent results.

For completeness in appendix \ref{toric}, we give a brief
introduction to some basic rules of toric geometry. We describe
there how one can identify the various compact divisors from the
toric data as well as the general rules for computing the triple
intersections of divisors, which will be used in later sections.

\subsection{Working with local Chern characters}

After this prelude, we now move to the more practical aspects of
working with local Chern characters. It was realised in
\cite{DelaOssa:2001xk} that the use of local Chern characters is
necessary even for D-branes associated with compact submanifolds
such as the fractional zero-branes. This is also true for our
application where we consider fractional two-branes.

In this paper will be dealing exclusively non-compact manifolds
$X$ which are the crepant resolution of singular spaces
$\BC^3/\Gamma$, for some abelian discrete group $\Gamma\subset
SU(3)$. Let $D_1$, $D_2$ and $D_3$ denote the non-compact
divisors and $D_4$, $D_5$, \ldots denote the compact divisors
that are added to resolve the singular space $\BC^3/\Gamma$.
These divisors thus correspond to four-cycles of
$X$.\footnote{Poincar\'e duality provides an isomorphism between
$H_4(X)$ and $H^2(X)$. Thus, when we write $D_i$ in the Chern
character, we take $D_i\in H^2(X)$.} The double intersection of
non-compact divisors give non-compact two-cycles and the other
double intersections are compact two-cycles of $X$.  If the
triple intersection involves at least one compact divisor, then
we get points with compact moduli spaces. Triple intersections
that do not involve at least one compact divisor are set to
zero.\footnote{ Mathematically speaking, it is well-known that
the only non-zero triple intersections are the ones that involve
at least one compact divisor. For an illustrative discussion of a
closely related result in a related context see for instance the
discussion in sec. 9, ch.2 of Iversen's text\cite{iversen}.}

As already mentioned, one has linear equivalences amongst the
divisors which lead to equivalences amongst the four- and
two-cycles as well. In the non-compact situation, one has to be
careful in using linear equivalences involving the non-compact
and compact divisors. For instance, consider $X$ to be the
resolution of $\BC^3/\BZ_3$. The manifold $X$ has one compact
divisor, $D_4=\BP^2$.  The linear equivalences are:
$$
D_1 \sim D_2 \sim D_3\ ,\ \textrm{and}\quad D_1+D_2+D_3+D_4 \sim 0\ .
$$
The second linear equivalence is valid {\it only} in the presence of
a compact divisor. By this we mean that
$$
 D_4 \cdot (D_1+D_2+D_3+D_4) \sim 0\quad \textrm{but}\quad 
 D_1 \cdot (D_1+D_2+D_3+D_4) \nsim 0\ .
$$
With this caveat in mind, we can use the linear equivalences to simplify
expressions.

The obvious inclusion maps for a (compact) divisor $j:\ D
\rightarrow X$ can be used to push-forward the Chern characters
of vector bundles on $D$ to {\it local} Chern characters of
sheaves on $X$. For instance, the push-forward of the structure
sheaf ${\cal O}_{D}$ on $X$, denoted by $j_*({\cal O}_{D})$, is
given by the sequence
\begin{equation}
0\rightarrow {\cal O}_X(-D) \rightarrow {\cal O}_X \rightarrow j_*({\cal O}_{D}) \rightarrow 0
\end{equation}
which gives
\begin{equation}
\ch \big[j_*({\cal O}_{D})\big] = D - \frac12 D \cdot D + 
\frac16 D\cdot D\cdot D \ .
\end{equation}
The local Chern character of all line-bundles are obtained by
tensoring the above sequence suitably with an appropriate
line-bundle.  For vector bundles $E$ with support on a divisor
$D$ the push-forward can be worked out by considering a
resolution of $E$ in terms of line bundles and then by pushing
forward each term to $X$. The local Chern character of the
push-forward sheaf, $j_*(E)$, can then be computed in a
straightforward fashion. As an example, let $D$ denote the
compact divisor in the resolution of $\BC^3/\BZ_3$ and
$\Omega_{\BP^2}(1)$ the vector bundle given by the following
exact sequence (the Euler sequence)
\begin{equation}
0\rightarrow \Omega^1_{\BP^2}(1) \rightarrow {\cal O}_{\BP^2}^{\oplus 3} 
\rightarrow {\cal O}_{\BP^2}\rightarrow 0\ .
\label{omegasequence}
\end{equation}
The local Chern character of the push-forward of $\Omega^1_{\BP^2}(1)$  
is then given by
\begin{equation}
\ch\big[j_*(\Omega_{\BP^2}(1))\big] = 3\ \ch\big[j_*({\cal O}_{\BP^2})\big]
- \ch\big[j_*({\cal O}_{\BP^2}(1))\big]  \ .
\end{equation}

Before we write down an explicit form in specific cases for the
Chern character it is useful also to take into account some
simplification provided by linear equivalences and symmetry. Thus
for the Chern character of a sheaf $E$ on the blow-up of
$\BC^3/\BZ_3$, we can write the following ansatz:
\begin{equation}
\label{ansatz1}
{\rm ch}(E)= a^\prime_1 + a_2D_4 + a_2^{\prime}D_1 +
 a_3D_1\cdot D_4 + a_3^{\prime}D_1\cdot D_2 + a_4p\ ,
\end{equation}
where the prime indicates terms involving only non-compact
divisors. Note that we have used the linear equivalence among the
$D_i (i\neq 4)$, but have not used it for $D_4$. We remind the
reader that the allowed triple intersection terms should have at
least one $D_4$ in them. To save on notation we have clubbed all
triple intersection terms together and re-written the term as a
single coefficient times the class of a point $p$.

Obviously for the toric description of the blow-up of other
$\BC^3/\BZ_n$ orbifolds the form of ch($E$) will be different
depending on the structure of the appropriate fan and the
corresponding compact and non-compact divisors. We write them
down for specific cases in the sequel.

\section{Local Chern characters of the fractional two-branes: $\BC^3/\BZ_3$}
\subsection{General method}
\label{genmethod}

We now discuss the general method that we use to compute the local Chern
character for the fractional two-branes. 
Unlike the fractional zero-branes where one needs to use the McKay
correspondence to obtain the Chern character, for the fractional two-branes
additional information is given by the open-string Witten index for
strings connecting them to fractional zero-branes
and as we will demonstrate will prove sufficient to fix the local Chern
character of the fractional two-branes.
The inputs that are used to compute the Chern character of the
fractional two-branes are the following.
\begin{itemize}
\item The intersection matrix (this is called ${\cal I}^{0,2}$ in the
sequel) which encodes the open-string 
Witten indices between the various fractional zero- and two-branes. This is
taken from the CFT computation since it is independent of K\"ahler
moduli.
\item An ansatz for the local Chern character analogous to Eq.
(\ref{ansatz1}) that takes into account  linear equivalences
among divisors.
\item The non-compact terms in the local Chern character are identical
for all fractional two-branes and equals $D_i\cdot D_j$ (for some $i,j$)
when the fractional two-brane is given by $\phi_i=\phi_j=0$. This
corresponds to a specific choice for the coefficients associated
with the non-compact divisors (we indicate these coefficients with
a prime in our ansatz). This is justified later.
\end{itemize}
With these inputs, the local Chern character is {\em uniquely} fixed up to the
class of a point. 
We will show that these Chern characters that we calculate provide a
justification for the imprecise and heuristic way of accounting for the
fractional charge that we had given in paper I.

We will now present two arguments, one geometric and the other
a worldsheet one, to show that the non-compact
contribution to the Chern character of the fractional two-brane is
given by $D_i\cdot D_j$.  The geometric argument is as follows. As a
concrete example, consider
the non-compact terms in the ansatz for $\BC^3/\BZ_3$ as given in
Eq. (\ref{ansatz1}). A non-vanishing value for $a_1^\prime$ corresponds
to a (non-compact) $D6-brane$ wrapping the resolution of  $\BC^3/\BZ_3$  while a
non-vanishing value for $a_2^\prime$ corresponds to a (non-compact) $D4$-brane
wrapping the four-cycle given by say, $\phi_1=0$. Clearly, this cannot
be the case since the large-volume limit corresponds to blowing up
compact four-cycles. Thus we conclude $a_1^\prime=a_2^\prime=0$.
This argument is valid for other examples as well. For a fractional
two-brane given by $\phi_i=\phi_j=0$, again one can rule out all
contributions other than $D_i\cdot D_j$. The second argument
is a worldsheet one. Recall that the resolution of the orbifold
singularity arise from closed-string moduli that appear in the twisted
sector in the CFT.
At the CFT point, the D-branes
are described by boundary states which can be schematically
written as
\begin{equation}
|B\rangle = |B\rangle_{\textrm{untwisted}}+ |B\rangle_{\textrm{twisted}}\ .
\label{boundarystate1}
\end{equation}
where we have explicitly separated contributions from the
untwisted sector and the twisted sector. The
$|B\rangle_{\textrm{untwisted}}$ does not couple to the K\"ahler
moduli. In particular, the one-point function on a disk of the
corresponding vertex operator obtains a vanishing contribution
from the untwisted sector. This also implies that the separation
in Eq. (\ref{boundarystate1}) holds even after resolving the
singularity and in particular, in the large volume limit. 
Thus, one expects
\begin{equation}
|B\rangle^{\textrm{large volume}} = |B\rangle_{\textrm{untwisted}}
+ |B\rangle_{\textrm{twisted}}^{\textrm{deformed}}\ ,
\label{boundarystate2}
\end{equation}
where the twisted sector boundary state is deformed by the K\"ahler
moduli while the untwisted sector boundary state is identical to the
CFT one.  The
non-compact terms arise solely from
$|B\rangle_{\textrm{untwisted}}$ and are identical for all
fractional two-branes and matches the expectation that it be
$D_i\cdot D_j$ for a fractional brane given by the boundary condition
$\phi_i=\phi_j=0$ at the CFT end. This is also consistent with the
geometrical picture of these fractional two-branes as extended
objects not localised at the singularity.

\subsection{Toric geometry of  $\BC^3/\BZ_3$}

The simplest orbifold to consider is $\BC^3/\BZ_3$ (and its
unique crepant resolution) with $\BZ_3$ action
$\frac13(1,1,1)$.\footnote{ The notation used here follows the
one used in paper I.} The resolution of the orbifold requires the
blowing up of the singular point at the origin to a $\BP^2$. The
toric data associated with the orbifold is given by three vectors
(see Appendix B and ref. \cite{Aspinwall:1994ev} for a review)
\begin{equation}
v_1=\begin{pmatrix}  1 \\ 0 \\ 0 \end{pmatrix},\quad
v_2=\begin{pmatrix}  0 \\ 1 \\ 0 \end{pmatrix},\quad
v_3=\begin{pmatrix}  -1\\ -1 \\ 3 \end{pmatrix}.
\end{equation}
The crepant resolution of the orbifold is given by the addition of
one vector:
\begin{equation}
v_4=\begin{pmatrix}  0 \\ 0 \\ 1 \end{pmatrix}.
\end{equation}
The vector $v_4$ is associated with a compact divisor $D_4=\BP^2$.  The
four vectors are not independent and satisfy a relation, which we write
as
$$
\sum_{i=1}^5 Q_i\ v_i =0 \ ,\quad a=1,2
$$
with 
$$
Q_i = 
\begin{pmatrix}  1 & 1 & 1 & -3 
\end{pmatrix} 
$$
This toric data is represented by the figure given below, in which the
various cones have been labelled as well. 
%\begin{figure}[ht]
%\begin{center}
\FIGURE{\includegraphics[width=3.5in]{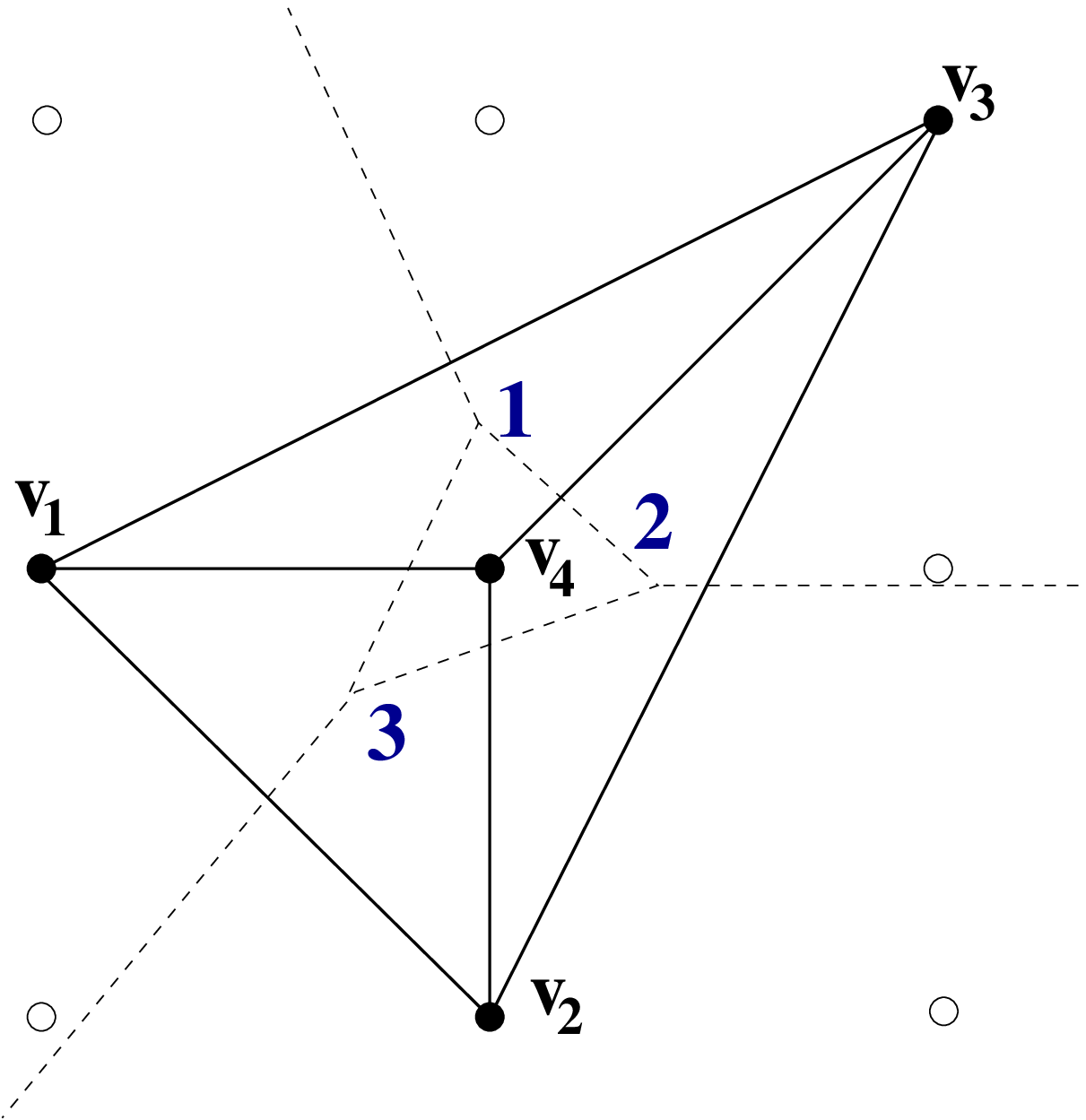}
\caption{Toric diagram for $\BC^3/\BZ_3$. The dotted lines indicated the
dual polytope.}
\label{fig1}
}
%\end{center}
%\end{figure}

The toric data can be naturally interpreted in terms of the
Gauged Linear Sigma Model (GLSM)\cite{Witten:1993yc}. The GLSM
associated with this toric data consists of four fields $\phi_i$
(one for each vector $v_i$) and one $U(1)$ with charge vectors
$Q_i$. The D-term equations are
\begin{eqnarray}
|\phi_1|^2 + |\phi_2|^2 + |\phi_3|^2 -3 |\phi_4|^2 = r 
\label{zthreedterm}
\end{eqnarray}
From the above D-term condition, we see that for $r\gg 0$ and
$\phi_4=0$, we have a $\BP^2$ with homogeneous coordinates
$\phi_1$, $\phi_2$ and $\phi_3$.  The orbifold limit is obtained
when $r\ll 0$. Here $\phi_4$ necessarily has a non-zero vacuum
expectation value $=\sqrt{|r|/3}$ and the $U(1)$ is broken to a
$\BZ_3$ with an action of $\frac13(111)$ on $\phi_1$, $\phi_2$
and $\phi_3$ respectively. The divisors $D_i$ are associated with
the four-cycles given by $\phi_i=0$.

\subsection{Triple Intersections}

The linear equivalences among the divisors are 
$$
D_{1}{\sim}D_{2}{\sim}D_{3}\, \textrm{ and }
D_{1}+D_{2}+D_{3}+D_{4}{\sim}0  \ .
$$
These equivalences are valid in the presence of a compact divisor.
Intersections of the compact divisors among themselves are
\begin{eqnarray}
&&D_{4}^3=9,\ D_4^2 \cdot D_1=-3,\ D_4 \cdot D_1^2 =1, \\
&&D_{4} \cdot D_{1}=h ,\ D_4^2  = -3h \ .
\end{eqnarray}
From the above intersections we can write down the 
intersections of the compact and non-compact divisors with  $h$ 
\begin{equation}
D_{4} \cdot  h=-3, \
D_{1} \cdot  h=1\ . 
\label{trieight}
\end{equation}

\subsection{Fractional zero-branes}

At the orbifold point, we impose Dirichlet boundary conditions,
$\phi_i=0$, $i=1,2,3$. We get {\em three} fractional boundary
states associated with these boundary conditions with a $\BZ_3$
which cyclically permutes them. We label the analytic
continuation of these D-branes to large-volume by $S_a^{(0)}$
($a=1,2,3$). The open-string Witten index is invariant under this
analytic continuation and provides an important input in our
analysis\cite{Douglas:1999hq}. At large-volume, it becomes the
intersection form which we denote by $\langle E,F \rangle$ for
two sheaves $E$ and $F$. It is the defined by
\begin{equation}
\langle E,F \rangle = \int_X {\rm ch}(E^*)\ {\rm ch}(F)\  \textrm{Td}(X)\ .
\end{equation}
Define the matrix:
\begin{equation}
{\cal I}^{0,0}_{a,b}=\langle S^{(0)}_{a}, S^{(0)}_{b}\rangle \ .
\end{equation}
Their intersection form computed as the open-string Witten index in
the CFT is found to be
\begin{equation}
{\cal I}^{0,0} = -(1-g)^3\ ,
\end{equation}
where $g$, the generator of the quantum $\BZ_3$ at the orbifold
point is given by the shift matrix
\begin{equation}
g =  \begin{pmatrix}
      0& 1& 0\\ 0& 0& 1 \\ 1& 0& 0
     \end{pmatrix}\ .
\end{equation}

The Chern classes for the fractional zero-branes in this example
is well known in the literature and were first determined in
\cite{Diaconescu:1999dt}.  The Chern classes for the fractional
zero-branes are known to be
\begin{eqnarray}
\ch\big[ S^{(0)}_{0}\big] &=& D_{4}+ (3/2)h +(3/2)p =\ch\big[j_*({\cal O}_{\BP^2})\big] \ ,
\nonumber \\
\ch\big[ S^{(0)}_{1}\big] &=& -2D_{4}  -2h -p  =-\ch\big[j_*(\Omega_{\BP^2}(1))\big]\ , \\
\ch\big[ S^{(0)}_{2}\big] &=& D_{4} +h/2 +p/2 =\ch\big[j_*({\cal O}_{\BP^2}(-1))\big]\ ,\nonumber
\end{eqnarray}
where $j:D_4 \rightarrow X$. On can independently verify that the
Chern classes written above are compatible with the CFT data.

\subsection{Fractional two-branes}

The fractional two-branes are obtained in the CFT by imposing
Dirichlet boundary conditions $\phi_1=\phi_2=0$ and imposing a
Neumann boundary condition on $\phi_3$. Using the 
general method in sec. \ref{genmethod},
 we are ready to determine the Chern characters of the
large-volume analogues of the fractional two-branes. Define the
matrices:
\begin{eqnarray}
{\cal I}^{0,2}_{a,b}&=&\langle S^{(0)}_{a}, S^{(2)}_{b}\rangle\ , \nonumber \\
{\cal I}^{2,2}_{a,b}&=&\langle S^{(2)}_{a}, S^{(2)}_{b}\rangle\ ,
\end{eqnarray}
where $S^{(0)}_a$ are the large volume analogues of the fractional
zero-branes and the $S^{(2)}_b$ are the corresponding objects for the
fractional two-branes. 

From the CFT computations for the $\BC^3/\BZ_3$ orbifold in I, we
know that
\begin{equation}
{\cal I}^{0,2}= -(1-g)^{2}\quad  \textrm{and} \quad
{\cal I}^{2,2}= g(1-g) \ .
\label{cftdata2}
\end{equation}
It is now simple to insert ansatz (\ref{ansatz1}) that we had
written down for the Chern character of the $S^{(2)}_i$ above and
try to solve for the coefficients using the CFT data given in Eq.  
(\ref{cftdata2}). The non-compact part of the Chern character is
taken to be $D_1\cdot D_2$, i.e., $a_1^\prime=a_2^\prime=0$ and
$a_3^\prime=1$ for all three fractional two-branes. The Chern
classes of the fractional two-branes as obtained from the
orbifold intersection from is
\begin{eqnarray}
\ch(S^{(2)}_{0})&=&D_{4}+(3/2)h+D_{1}{\cdot}D_{2} + a_4~p\ ,\nonumber \\
\label{coeffs}
\ch(S^{(2)}_{1})&=&-D_{4}-(1/2)h+D_{1}{\cdot}D_{2}+ b_4~p \ ,\\
\ch(S^{(2)}_{2})&=&D_{1}{\cdot}D_{2} + c_4~p\ ,\nonumber
\end{eqnarray}
where $a_4,b_4,c_4$ are the coefficients that are {\em not} fixed
by the intersection numbers. Again, these choices are compatible
with the second intersection form given in Eq. (\ref{cftdata2}).
Note that ${\cal I}^{2,2}$ was not used in fixing the
coefficients appearing in the ansatz (\ref{ansatz1}) and thus it
can be used as an additional check.

\subsection{Geometry of the fractional two-branes}

We can now see that these Chern characters provide some precise
justification for the heuristic constructions of Paper I where we had
constructed the large volume analogues of the
fractional two-branes using some exact sequences (in particular,
see  Sec. 4.5, Eq.  (4.20)-(4.22)). These sequences
however were imprecisely defined, mathematically speaking,
since we introduced fractional Chern classes for these objects.
The meaning of these sequences now becomes clearer. The objects
indicated by these sequences will be interpreted as contributing to
the compact part of the Chern character, thus retaining their
original mathematical meaning, while the fractional part will
contribute precisely to the non-compact $D_1\cdot D_2$ term of the Chern
character. This (implicitly additive) separation is allowed since the
Chern character is also additive for direct sums. We can rewrite
Eq. (\ref{trieight}) as $D_4 \cdot D_4 \cdot D_2=-3$ and $D_1 \cdot
D_4\cdot D_2=1$ utilising linear equivalence. In the background of
a compact divisor, namely $D_4$, this reflects the fact that the 
non-compact contribution, $D_1\cdot D_2$ accounts for exactly $1/3$ of
the unit charge given by the $D_4\cdot D_2$ contribution.
Thus using the Eq. (4.26)-(4.28) of paper I we obtain the
following\footnote{An independent explanation is also given below.}:
\begin{eqnarray}
\ch\big[S_0^{(2)}\big] &=& \ch\big[j_*({\cal O}_{\BP^2}) \big] + D_1\cdot D_2 \ ,
\nonumber \\
\ch\big[S_1^{(2)}\big] &=& \ch\big[j_*({\cal O}_{\BP^2}(-1) \big] 
+ D_1\cdot D_2\ , \\
\ch\big[S_2^{(2)}\big] &=& D_1\cdot D_2 \ .\nonumber
\end{eqnarray}
We can compute the Chern character of the pieces $\ch[j_*(..)]$
using the techniques discussed in Sec. 2
and verify that indeed we reproduce the Chern characters
computed in the previous subsection in Eq. (\ref{coeffs}).
Thus the apparently imprecise sequences of paper I are indeed
meaningful provided we exercise some care in interpretation.
 
We can now of course recall the physics side of the story from paper
I in the language of what we had referred to as Higgs and Coulomb
branes and their construction in terms of the bulk fermion degrees
of freedom that survive on the boundary of the world-sheet.
In order to do this, we first separate the
contributions when $\phi_3=0$ from those that are present when
$\phi_3\neq0$. At the orbifold point, the first contribution is
localised at the singularity. This contribution is similar to
that of fractional zero-branes on a $\BC^2/\BZ_3$ orbifold (where
the coordinates of $\BC^2$ are $\phi_1$ and $\phi_2$). Due to
this similarity, following Martinec and Moore\cite{MM}, we call
the contribution when $\phi_3=0$ as the {\em Higgs branch} and
the contribution when $\phi_3\neq0$ the {\em Coulomb branch}. As
we will see, some of the fractional two-branes have both branches
-- we call them the {\em Higgs branes} while some only have a
Coulomb branch -- we call them the {\em Coulomb branes}.

We will now use a method that was alluded to but not used in paper I
to extract the Higgs branch contributions of the fractional two branes.  
Now, there are only two fermions $\xi_1$ and $\xi_2$ coming from
the Dirichlet boundary conditions\footnote{For Dirichlet boundary
conditions preserving B-type supersymmetry, the $\xi$'s are the
linear combination {\it not} set to zero. Further, these are the
``observables'' in the topological B-model. See paper I for a more
detailed discussion.} since the fermionic partners of the Neumann
scalars do not contribute as suggested in paper I.  The compact part of
the fractional two-branes are in one-to-one correspondence with
the states given in the table below (the vacuum $|0\rangle$
satisfies $\bar{\xi}_i |0\rangle =0$ for $i=1,2$)  subject to
the gaugino constraint being satisfied:
\begin{equation}
\phi_1 \bar{\xi}_{1} + \phi_2\bar{\xi}_{2} =
0\ .
\label{gauginoc3z3}
\end{equation}

\begin{center}
\begin{tabular}{c|c|c|c}
\hline
Label &$S^{(2)}_0$ & $S^{(2)}_1$ & $S^{(2)}_2$  \\[3pt] \hline
$U(1)$ charge & 0 & 1 & 2  \\[3pt] \hline
State &$|0\rangle$&
$\xi_i|0\rangle$&
$\xi_1 \xi_2 |0\rangle$ \\[3pt] \hline 
Higgs branch & $D_4$ & $D_4$  & --- 
\\[3pt]
\hline
\end{tabular}
\end{center}
In the above table, we indicate based on the detailed discussion
below, the divisor on which the gaugino constraint can be
satisfied when $\phi_3=0$. As explained earlier, we will
distinguish the contribution when $\phi_3=0$ and when $\phi_3\neq
0$ -- both possibilities are permitted by the Neumann boundary
condition.  In the Higgs branch, at large volume,
$\phi_1=\phi_2=0$ is not allowed. This is however allowed in the
Coulomb branch where $\phi_3\neq0$. Thus the Coulomb branch
contribution to the Chern classes of the fractional two-branes
contain $D_1\cdot D_2$. This is consistent with the Chern
character that we obtained for $S_2^{(2)}$ in Eq. (\ref{coeffs}).
\begin{enumerate}
\item[$S^{(2)}_1$]   The gaugino constraint implies that we have a rank one
bundle in the Higgs branch which can be identified with 
$j_*({\cal O}_{\BP^2}(-1))$.
\item[$S^{(2)}_2$] This is a Coulomb brane since the gaugino constraint
cannot be satisfied when $\phi_3=0$.
\item[$S^{(2)}_0$]  This is a line bundle which can be identified with 
$j_*({\cal O}_{\BP^2})$.
\end{enumerate}
In this example, the Coulomb branch is identical
to the non-compact contribution to the Chern class and is
identical for all three fractional two-branes. 
As we will see, in the next example, the
Coulomb branch is {\it not} the same as the non-compact term
though it contains it. In the more
intricate cases we will need both the precise Chern character
computation and the physics construction of the boundary states using
the fermions and their interpretation to pin down the objects
corresponding to the large-volume analogues of all the fractional
two-brane states

\subsection{A change of basis}

As we just saw, the Higgs branes have two kinds of contributions,
one from the Higgs branch and the other from the Coulomb branch.
We will now exhibit an {\it integral} change of basis which
removes the Coulomb branch from the Higgs branes. Let
$\mathbf{\hat{S}^{(2)}}=(\hat{S}^{(2)}_0,\hat{S}^{(2)}_1,\hat{S}^{(2)}_2)^{\rm
T} $ represent the new basis and $\mathbf{{S}^{(2)}}$ the
original basis of fractional two-branes. Then,
\begin{equation}
\mathbf{\hat{S}^{(2)}} =\begin{pmatrix} 1 & 0 & -1 \\ 0 & 1 & -1 \\ 0 & 0 & 1 \end{pmatrix}
\ \mathbf{{S}^{(2)}}\ .
\end{equation}
is the required change of basis. Note that it is an upper-triangular matrix. 

\section{The $\BC^3/\BZ_5$ example}

We will consider the example of the orbifold $\BC^3/\BZ_5$ with the
action $\frac15(1,1,3)$. As discussed in Appendix \ref{toric},
the resolution of this requires the blowing up
of a $\BP^2$ and a Hirzebruch surface $\BF_3$ (which is a $\BP^1$
fibration over $\BP^1$)\cite{DelaOssa:2001xk,Mukhopadhyay:2001sr}.  The
two exceptional divisors intersect along a curve which is a hyperplane
on the $\BP^2$. The toric data associated with the orbifold is given by
three vectors
\begin{equation}
v_1=\begin{pmatrix}  1 \\ 0 \\ 0 \end{pmatrix},\quad
v_2=\begin{pmatrix}  -1 \\ -3 \\ 5 \end{pmatrix},\quad
v_3=\begin{pmatrix}  0\\ 1 \\ 0 \end{pmatrix}.
\end{equation}
The unique crepant resolution of the orbifold is given by the addition of
two vectors:
\begin{equation}
v_4=\begin{pmatrix}  0 \\ -1 \\ 2 \end{pmatrix},\quad
v_5=\begin{pmatrix}  0 \\ 0 \\ 1 \end{pmatrix}.
\end{equation}
As explained in Appendix \ref{toric},
the vector $v_4$ is associated with the $\BP^2$ and $v_5$ with the
Hirzebruch surface $\BF_3$. The five vectors are not independent
and satisfy two relations, which we write as
$$
\sum_{i=1}^5 Q_i^A\ v_i =0 \ ,\quad A=1,2
$$
with 
$$
Q_i^A = 
\begin{pmatrix}  1 & 1 & 0 & -3 & 1 \\ 
                 0 & 0 & 1 & 1 &-2 
\end{pmatrix} \ .
$$
In figure \ref{fig2}, the toric data 
is represented by the following projection on a
two-dimensional plane.
%\begin{figure}[ht]
%\begin{center}
\FIGURE{\includegraphics[width=3.5in]{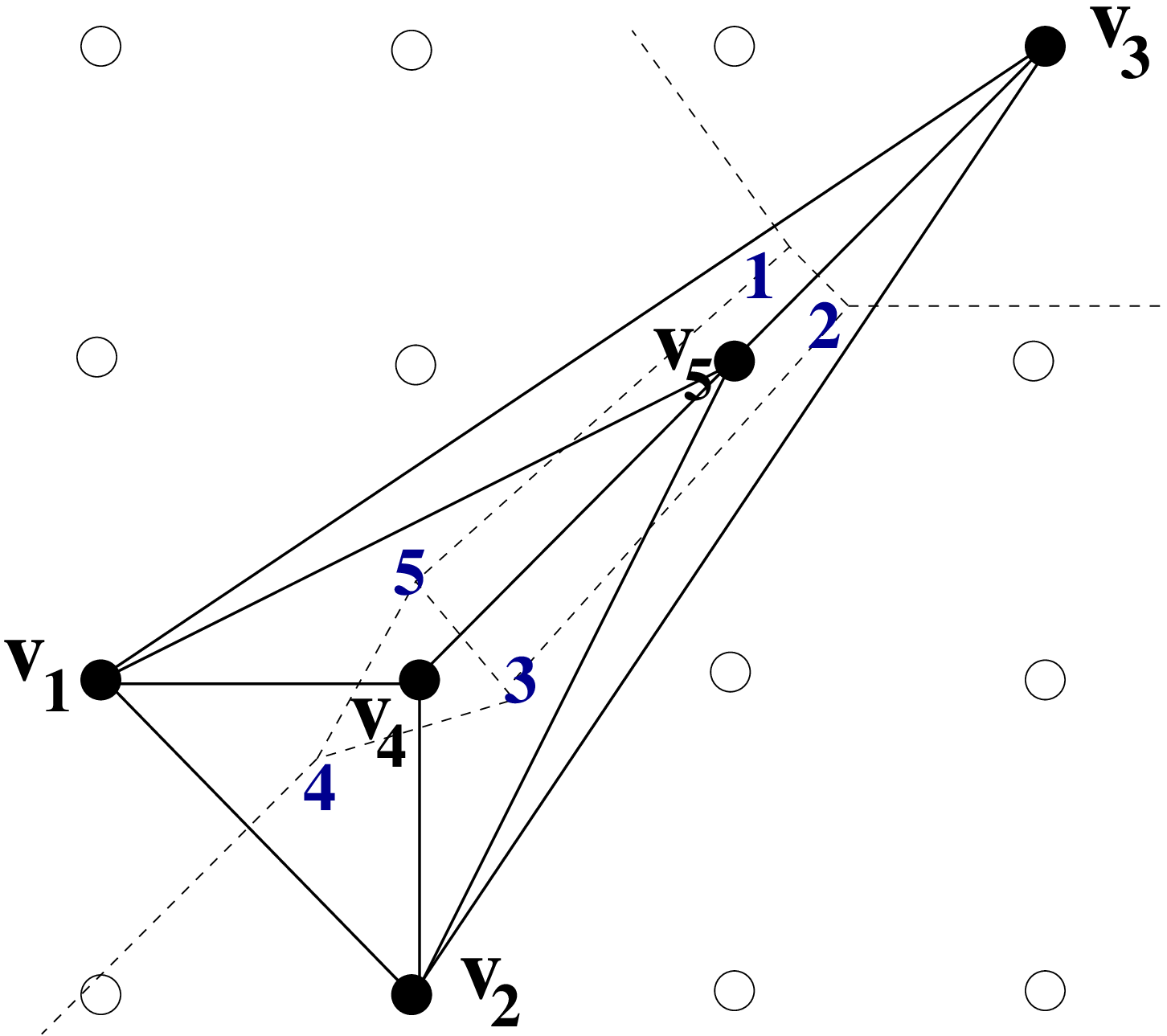}
%\end{center}
\caption{Toric diagram for $\BC^3/\BZ_5$}
\label{fig2}
}
%\end{figure}

The GLSM associated with this toric data consists of five fields
$\phi_i$ (one for each vector $v_i$) and two $U(1)$'s (one for each
relation) with charge vectors $Q^A_i$. The D-term equations are
\begin{eqnarray}
|\phi_1|^2 + |\phi_2|^2 + |\phi_5|^2 -3 |\phi_4|^2 = r_1\ , \nonumber \\
|\phi_3|^2 + |\phi_4|^2  -2 |\phi_5|^2 = r_2 \ .
\label{dtermsc3z5}
\end{eqnarray}
From the first D-term condition, we see that for $r_1\gg 0$ and
$\phi_4=0$, we have a $\BP^2$ with homogeneous coordinates $\phi_1$,
$\phi_2$ and $\phi_5$.  The base of $\BF_3$ is the $\BP^1$ is given by
the hypersurface $\phi_5=0$ in the $\BP^2$ and thus has homogeneous
coordinates $\phi_1$ and $\phi_2$. The second D-term for $r_2\gg0$ and
$\phi_5=0$, gives a $\BP^1$ with homogeneous coordinates $\phi_3$ and
$\phi_4$. This $\BP^1$ is the fibre of $\BF_3$. 

The orbifold limit is obtained by first considering a particular linear
combination of the above two D-term conditions, i.e., the one associated
with $Q_i\equiv (Q_i^1 + 3 Q_i^2)$:
\begin{equation}
|\phi_1|^2 + |\phi_2|^2 + 3 |\phi_3|^2 - 5 |\phi_5|^2 = r_1 + 3 r_2\ .
\label{orbifolddterm}
\end{equation}
In the limit $(r_1+3r_2)\ll 0$, $\phi_5$ necessarily has a non-zero vev
$=\sqrt{|r_1+3r_2|/5}$ and the associated $U(1)$ is broken to a $\BZ_5$
with an action of $\frac15(113)$ on $\phi_1$, $\phi_2$ and $\phi_3$
respectively.

\subsection{Triple Intersections}

The linear equivalences among the divisors are 
$$
D_{1}{\sim}D_{2}{\sim}D_{5}+2D_{3}\, \textrm{ and }
D_{1}+D_{2}+D_{3}+D_{4}+D_{5}{\sim}0   \ .
$$
These equivalences are valid in the presence of a compact divisor.
Intersections of the compact divisors among themselves are
\begin{equation}
D_{4}^3=9,\ D_{4}^2{\cdot}D_{5}=-3, \
D_{4}{\cdot}D_{5}^2=1, \
D_{5}^3=8\ .
\end{equation}
Intersections of the compact divisors with the non-compact divisors are  
\begin{eqnarray}
D_{4}^{2}\cdot D_{1}=-3,\ D_{4}\cdot D_{5}\cdot D_{1}=1, \
D_{5}^{2}\cdot D_{1}=-2, \
D_{5}^{2}\cdot D_{3}=-5\ , \nonumber \\
D_{1}^{2}\cdot D_{4}=1,\ D_{5}\cdot D_{1}^{2}=0, \
D_{5} \cdot D_{1} \cdot D_{3}=1, \
D_{3}^{2}{\cdot}D_{5}=3\ .
\end{eqnarray}
We also have
\begin{equation}
D_{4} \cdot D_{5}= D_{4} \cdot D_{1}=h, \
D_{4} \cdot D_{3}=0, \ D_{5} \cdot D_{1}=f, \
D_{3} \cdot D_{5}=h+3f\ .
\end{equation}
where $f$ is the $\BP^1$ fibre of $\BF_{3}$ and $h$ is the
hyperplane in $\BP_{2}$.  The self intersections of $D_{4}$ and
$D_{5}$ are
$$
D^{2}_{4}=-3h, \
D^{2}_{5}=-2h-5f\ .
$$
From the above intersections we can write down the 
intersections of the compact and non-compact divisors with  $h$ and $f$
\begin{eqnarray}
D_{4}{\cdot}h=-3, \
D_{4}{\cdot}f=1, \ D_{5}{\cdot}h=1, \
D_{5}{\cdot}f=-2\ ,
\nonumber \\
D_{1}{\cdot}h=1, \
D_{1}{\cdot}f=0, \ D_{3}{\cdot}h=0, \
D_{3}{\cdot}f=1\ .
\end{eqnarray}

\subsection{Fractional zero-branes}

At the orbifold point, we impose Dirichlet boundary conditions,
$\phi_i=0$, $i=1,2,3$. We get {\em five} fractional boundary
states associated with these boundary conditions with a $\BZ_5$
which cyclically permutes them. Their intersection form is
\begin{equation}
{\cal I}^{0,0} = -(1-g)^2(1-g^3)\ ,
\end{equation}
where $g$ denotes the $5\times 5$ shift-matrix that generates the
$\BZ_5$. One can choose an ansatz analogous to the one in
(\ref{ansatz1}) and determine the Chern characters of the
fractional branes using the above intersection form. The relevant
ansatz is
\begin{equation}
\label{ansatz2}
\ch(E)=a_1' + a_2 D_4 + a_3 D_5 + a_2' D_1 + a_3' D_3 + a_4 h + a_5 f + a_4' D_1^2 + a_5' D_1\cdot D_3 + a_6 p\ .
\end{equation} 
Of course, the compact nature of the fractional zero-branes
implies that all the non-compact pieces (indicated by adding a
prime to the coefficient) vanish for fractional zero-branes. The
fractional zero-branes are known to have the following local
Chern character
\begin{eqnarray}
 \ch\big[S^{(0)}_{0}\big] &=& D_{4} + D_{5} +(3/2)h +(5/2)f \ , \nonumber \\
 \ch\big[S^{(0)}_{1}\big] &=& -2D_{4} -D_{5} -2h -(3/2)f \ , \nonumber \\
 \ch\big[S^{(0)}_{2}\big] &=& D_{4} +h/2 \ , \\
 \ch\big[S^{(0)}_{3}\big] &=& -D_{5} -(5/2)f\ , \nonumber  \\
 \ch\big[S^{(0)}_{4}\big] &=& D_{5} +(3/2)f\ . \nonumber
\end{eqnarray}
These were first obtained in
\cite{Mukhopadhyay:2001sr,DelaOssa:2001xk} using the McKay
correspondence which we discuss in section 5 to obtain the Chern
characters of the fractional zero-branes. This is done by first
obtaining the tautological bundles and then computing their
duals. The Chern character is clearly compatible with
Eq.(\ref{ansatz2}) and one sees that the coefficients of the
non-compact terms vanish as expected.

As this has not been discussed earlier in the literature, we 
will now write out concrete objects which will correspond to
specific choices for the coefficient of the class of a point
using the physical method proposed in paper I. The gaugino
constraint (for the vector multiplet associated with the D-term
in Eq. (\ref{orbifolddterm})) is
\begin{equation}
\phi_1 \bar{\xi}_{1} + \phi_2\bar{\xi}_{2} + 3 \phi_3\bar{\xi}_{3} =
0\ .
\label{gauginoc3z5}
\end{equation}
The fractional zero-branes are in one-to-one correspondence with
the states\cite{bundlesglsm,mayr}: (the vacuum $|0\rangle$ satisfies
$\bar{\xi}_i |0\rangle =0$ and the index $a=1,2$)
\begin{center}
\begin{tabular}{c|c|c|c|c|c}
\hline
Label &$S^{(0)}_0$ & $S^{(0)}_1$ & $S^{(0)}_2$ & $S^{(0)}_3$ 
& $S^{(0)}_4$ \\[3pt] \hline
$U(1)$ charge & 0 & 1 & 2 & 3& 4 \\[3pt] \hline
State &$|0\rangle$&
$\xi_a|0\rangle$&
$\xi_1\xi_2|0\rangle$&
$\xi_3 |0\rangle$&
$\xi_a \xi_3 |0\rangle$ \\[3pt] \hline 
 & $D_4$ and $D_5$ & $D_4$  and $D_5$ & $D_4$ & $D_5$& $D_5$
\\[3pt]
\hline
\end{tabular}
\end{center}
subject to the gaugino constraint Eq. (\ref{gauginoc3z5}) being
satisfied. In the third row of the above table, we indicate the divisors
on which this is possible based on the following considerations.
Let us consider the various states and identify the
corresponding coherent sheaves.
\begin{enumerate}
\item[$S^{(0)}_1$] The gaugino constraint is trivially satisfied
when both $\phi_1=\phi_2=0$. Thus, the sheaf has rank two when
$\phi_1=\phi_2=0$ and rank one when either $\phi_1\neq0$ and/or
$\phi_2\neq0$. By studying the two D-terms given in Eq.
(\ref{dtermsc3z5}), we can see that the rank two condition is possible
(at large volume) only if $\phi_5\neq0$. Thus, the rank two part has
support on the compact divisor $D_4$ while the rank one part has
support on $D_5$. Thus, we expect the following to hold:
$$
S^{(0)}_1 = i_*(V) + j_*(W) \ .
$$
where $V$ is a rank-two bundle on $D_4=\BP^2$ and $W$ is a line-bundle
on $D_5=\BF_3$. In fact, one can identify $V$ with $\Omega_{\BP^2}(1)$
and $W$ with ${\cal O}_{D_5}(-D_1-D_4)$.
\item[$S^{(0)}_2$] The gaugino constraint holds only when $\phi_1=\phi_2=0$
and hence $S_2$ has support on $D_4$ where it is a line-bundle ${\cal
O}_{\BP^2}(-1)$.
\item[$S^{(0)}_3$] The gaugino constraint holds when $\phi_3=0$ which
requires $\phi_4\neq0$. Thus, $S^{(0)}_3$ is the push-forward of a
line-bundle on $D_5$. The line bundle can be identified with 
${\cal O}_{D_5}(-D_4)$.
\item[$S^{(0)}_4$] The gaugino constraint holds when $\phi_3=0$ and when
either $\phi_1\neq0$ or $\phi_2\neq0$. The D-term constraints imply
that $S_4$ is the push-forward of a line-bundle on $D_5$. The line bundle
can be identified with
${\cal O}_{D_5}(-D_1-D_4)$.

\item[$S^{(0)}_0$] The gaugino constraint trivially holds and hence $S_0$ is
the direct sum of the push-forward of line-bundles on $D_4$ and $D_5$.
The two line bundles are ${\cal O}_{\BP^2}$ and ${\cal O}_{D_5}(-D_4)$.
\end{enumerate}

We will now move on to the fractional two-branes next. There two
inequivalent types of fractional two-branes in this example -- we
can impose Neumann boundary condition on $\phi_1$ or $\phi_3$. We
will label them $A$ and $B$ respectively instead of the notation
$S^{(2)}$ used in the $\BC^3/\BZ_3$ example.

\subsection{Fractional two-branes -- Type I}

We will first consider the fractional two-branes obtained by imposing
a Neumann boundary condition on $\phi_1$ and Dirichlet boundary
conditions $\phi_2=\phi_3=0$.  The
master formula given in paper I provides us the intersection form
amongst the fractional two-branes as well as the intersection with the
fractional zero-branes. We obtain
\begin{eqnarray}
{\cal I}^{0,2} &=&-(1+g-g^{2}-g^{4})\ , \\
{\cal I}^{2,2} &=&-(g^{4}-g)\ .
\end{eqnarray}
The Chern classes of the fractional two-branes as obtained 
using the method discussed in sec. \ref{genmethod} are 
\begin{eqnarray}
\ch[A_{0}]&=&D_{4}+D_{5}+D_{2}{\cdot}D_{3}+(3/2)h+(5/2)f \ ,\nonumber \\
\ch[A_{1}]&=&-D_{4}+D_{2}{\cdot}D_{3}-(1/2)h+f\ , \nonumber \\
\ch[A_{2}]&=&D_{2}{\cdot}D_{3}+f \ ,\\
\ch[A_{3}]&=&-D_{5}+D_{2}{\cdot}D_{3}-(3/2)f\ , \nonumber \\
\ch[A_{4}]&=&D_2\cdot D_3\ .\nonumber  
\end{eqnarray}
The non-compact contribution is $D_2\cdot D_3$ as discussed in sec.
\ref{genmethod}.

We now proceed to obtain coherent sheaves  which reproduce 
the above Chern classes.
For the chosen boundary condition, there are only two fermions $\xi_2$ and $\xi_3$. The Higgs branch
of the fractional two-branes  are in one-to-one
correspondence with the states: (the vacuum $|0\rangle$ satisfies
$\bar{\xi}_i |0\rangle =0$ for $i=2,3$.) We will work out the component
corresponding to the Higgs branch i.e., when $\phi_1=0$. For $D_4$,
this picks out a $\BP^1\in D_4$ (with homogeneous coordinates $\phi_2$
and $\phi_5$) and a $\BP^1\in D_5$ (with homogeneous coordinates
$\phi_3$ and $\phi_4$).
\begin{center}
\begin{tabular}{c|c|c|c|c|c}
\hline
Label &$A_0$ & $A_1$ & $A_2$ & $A_3$ & $A_4$ \\[3pt] \hline
$U(1)$ charge & 0 & 1 & 2 & 3& 4 \\[3pt] \hline
State &$|0\rangle$&
$\xi_2|0\rangle$&
-- &
$\xi_3 |0\rangle$&
$\xi_2 \xi_3 |0\rangle$ \\[3pt] \hline 
Higgs branch & $D_4$ and $D_5$ & $D_4$  & --- & $D_5$& ---
\\[3pt]
\hline
\end{tabular}
\end{center}
subject to the gaugino constraint Eq. (\ref{gauginoc3z5}) being satisfied.

In the Higgs branch where $\phi_1=0$, at large volume, $\phi_2=\phi_3=0$
is not allowed and $\phi_2=\phi_5=0$ is also not allowed. These are
however allowed in the Coulomb branch where $\phi_1\neq0$. Thus the
Coulomb branch contributions to the Chern classes of the fractional
two-branes contain $D_2\cdot D_3$ and/or $D_2\cdot D_5 =f$. This is
consistent with the Chern classes that we obtain for $A_2$ and $A_4$ which
have a vanishing Higgs branch. Clearly, we see that the Coulomb branch 
involves a compact piece as well.
\begin{enumerate}
\item[$A_1$]    The gaugino constraint needs $\phi_2=0$. Thus, one
needs $\phi_5\neq0$. Thus, the fractional two-brane has support only on
$D_4$ and is the line-bundle ${\cal O}_{D_4}$.
\item[$A_2$] This is one of the Coulomb branes.
\item[$A_3$] This needs $\phi_3=0$. It has support on $D_5$ alone and
the Higgs branch is the line bundle ${\cal O}_{D_5}(-D_1-D_4)$.
\item[$A_4$] This needs $\phi_2=\phi_3=0$. So this is another Coulomb brane.
\item[$A_0$]  This has support on both $D_4$ and $D_5$
and the Higgs branch is the push-forward of two line bundles on these
divisors.
These  are ${\cal O}_{D_4}$ and ${\cal O}_{D_5}(-D_4)$ respectively.
\end{enumerate}
{\bf Comment:} $A_0$ $A_1$, $A_2$ are in some ways similar to fractional
two-branes on the resolution of $\BC^3/\BZ_3$.

\subsection{Fractional two-branes -- Type II}

We will consider the fractional two-branes obtained by imposing
Neumann boundary conditions on $\phi_3$ and $\phi_1=\phi_2=0$.
The master formula given in paper I provides us the intersection
form amongst the fractional two-branes as well as the
intersection with the fractional zero-branes. We obtain
\begin{eqnarray}
{\cal I}^{0,2}&=&-g^{2}(1-g)^{2}\ , \\
{\cal I}^{2,2}&=&g-g^{2}+g^{3}-g^{4}\ .
\end{eqnarray}
The Chern classes obtained from the intersection form at the orbifold
point are (ignoring the class of a point)
\begin{eqnarray}
\ch[B_{0}]&=&D_4+D_5+D_1\cdot D_2+(3/2)h+(5/2)f\ ,  \nonumber \\
\ch[B_{1}]&=&-D_4-D_5+D_1\cdot D_2-h/2-(3/2)f\ ,\nonumber \\
\ch[B_{2}]&=&D_1\cdot D_2 \ , \\
\ch[B_{3}]&=&D_4+D_1\cdot D_2+(3/2)h\ , \nonumber \\
\ch[B_{4}]&=&-D_4+D_1\cdot D_2-h/2\ . \nonumber 
\end{eqnarray}

For the given boundary condition, there are only two fermions
$\xi_1$ and $\xi_2$. The Higgs branch of the fractional
two-branes are in one-to-one correspondence with the states: (the
vacuum $|0\rangle$ satisfies
$\bar{\xi}_a |0\rangle =0$ for $a=1,2$.) 
\begin{center}
\begin{tabular}{c|c|c|c|c|c}
\hline
Label &$B_0$ & $B_1$ & $B_2$ & $B_3$ & $B_4$ \\[3pt] \hline
$U(1)$ charge & 0 & 1 & 2 & 3& 4 \\[3pt] \hline
State &$|0\rangle$&
$\xi_a|0\rangle$&
$\xi_1\xi_2|0\rangle$&
---&
--- \\[3pt] \hline 
Higgs branch & $D_4$ and $D_5$ & $D_5$ & ---  &  --- & ---
\\[3pt]
\hline
\end{tabular}
\end{center}
subject to the gaugino constraint Eq. (\ref{gauginoc3z5}) being satisfied.

The Higgs branch is when $\phi_3=0$. In this case, at large volume, it
is {\bf not} possible to have either $\phi_4=0$ or $\phi_1=\phi_2=0$.
This is possible in the Coulomb branch where $\phi_3\neq0$. Thus the
contributions of the Coulomb branes can arise these two sources.  The
associated Chern classes are $(D_4 + 3h/2 +3p/2)$ (from $\phi_4=0$) and
$D_1\cdot D_2$ (from $\phi_1=\phi_2=0$).
\begin{enumerate}
\item[$B_1$]  The contribution that arises on $D_5$ is as in the case
of $S^{(0)}_1$ and is the line-bundle ${\cal O}_{D_5}(-D_1-D_4)$. 
The $D_4$ appearing in the Chern character must come from the Coulomb 
branch since $\phi_4=0$ is not
allowed at large volume when $\phi_3=0$.
\item[$B_2$] This has support when $\phi_1=\phi_2=0$. In the Higgs
branch, $\phi_3=0$ and hence this is not allowed. Hence, this is a 
Coulomb brane.
\item[$B_3$] This is one of the Coulomb branes.
\item[$B_4$] This is one of the branes Coulomb branes.
\item[$B_0$]  The discussion is similar to $S^{(0)}_0$ and
the Higgs branch of the sheaf has support on both $D_4$ and $D_5$
and can be identified with 
the direct sum of the push-forward of the line-bundles
${\cal O}_{\BP^2}$ and ${\cal O}_{D_5}(-D_4)$.
\end{enumerate}

\subsection{A change of basis}

As we just saw, the Higgs branes have two kinds of contributions,
one from the Higgs branch and the other from the Coulomb branch.
We will now exhibit an {\it integral} change of basis which
removes the Coulomb branch from the Higgs branes. Let
$\hat{\mathbf{A}}=(\hat{A}_0,\ldots,\hat{A}_4)^{\rm T} $
represent the new basis for type I fractional two-branes and
$\hat{\mathbf{B}}$ the new basis for the type II fractional
two-branes. Then,
\begin{eqnarray}
\hat{\mathbf{A}} &=\begin{pmatrix} 1 & 0 & 0&0&-1 \\ 0 & 1 & -1 &0&0 \\ 0 & 0 & 1&0&0 \\
0&0&0&1&-1 \\ 0 & 0& 0& 0& 1 \end{pmatrix}
\ \mathbf{A} \ ,\\
\hat{\mathbf{B}} &=\begin{pmatrix} 1 & 0 & -1&0&0 \\ 0 & 1 & 0 &0&-1 \\ 0 & 0 & 1&0&0 \\
0&0&0&1&0 \\ 0 & 0& 0& 0& 1 \end{pmatrix}
\ \mathbf{B} \ .
\end{eqnarray}
are the required change of bases. It is of interest if this change of basis is related to a change of basis proposed by Moore and Parnachev in \cite{Moore:2004yt}.

\section{Quantum McKay correspondence}

The McKay correspondence \cite{Kay,reid,mkay} can be stated in
several different forms. We will consider the one due to Ito and
Nakajima\cite{Naka} where the McKay correspondence is presented
as a duality between two families of sheaves.  The first family
is given by the coherent sheaves associated with the fractional
zero-branes and the second one is associated with the so-called
$tautological$ bundles. The duality is stated
as\cite{DelaOssa:2001xk}
\begin{equation}
( S^{(0)}_{a},R^{b}_{(0)})_{X} \equiv \int_X \ch(R^{b}_{(0)})\ \ch(S^{(0)}_{a})\ {\rm Td}( X )=\delta^{b}_{a}\ ,
\end{equation}
where we remind the reader that $X$ is the crepant resolution of
the orbifold. Note that the above expression is not the
intersection form and hence we indicate the inner product by $( ,
)$ rather than $\langle , \rangle$. Inspired by this, a
generalisation called the quantum McKay correspondence was
proposed in I for the fractional $2p$ -branes and is stated as
the following duality:
\begin{equation}
( S^{(2p)}_{a},R^{b}_{(2p)})_{X} \equiv 
\int_X \ch(R^{b}_{(2p)})\ \ch(S^{(2p)}_{a})\ {\rm Td}(X) =\delta^{b}_{a}
\ .
\end{equation}
This seems to be related to a correspondence of Martinec and
Moore \cite{MM} in the context of non-supersymmetric orbifolds.
In the following we shall obtain the Chern characters of the
duals for the fractional two-branes in the two working examples
in the paper: $\BC^{3}/\BZ_{3}$, $\BC^{3}/\BZ_{5}$. As always,
the intersection numbers do not uniquely fix the Chern class due
to the linear equivalences among the divisors. Unlike the
fractional two-branes, we are unable to fix this ambiguity by
appealing to CFT.
 
\subsection{$\BC^{3}/\BZ_{3}$ orbifold}

The Chern characters for the R-sheaves corresponding to the Higgs branes are  
\begin{eqnarray}
\ch(R^{0}_{(2)})&=& D_{1} -\frac{D_{1}^{2}}{2}\ , \nonumber \\
\ch(R^{1}_{(2)})&=& D_{1} -\frac{3D_{1}^{2}}{2}\ , 
\end{eqnarray}
and for the R-sheaf dual to the Coulomb brane among the fractional two-branes,
the Chern character is 
\begin{equation}
\ch(R^{2}_{(2)})= D_{1} +D_{4} -\frac{2h}{3} \ .
\end{equation}

\subsection{$\BC^{3}/\BZ_{5}$ orbifold}

Recall that there are two types of fractional two-branes in this example.
We quote the duals for both types of fractional two-branes.

\subsection*{ Type I}

\noindent The Chern characters of the R-sheaves dual to the Higgs branes are :
\begin{eqnarray}
\ch(R^{0}_{(2)})&=& D_{1} -\frac{D_{1}^{2}}{2} \ ,\nonumber \\
\ch(R^{1}_{(2)})&=& D_{1} -\frac{3D_{1}^{2}}{2} \ ,\nonumber \\
\ch(R^{3}_{(2)})&=& D_{1} -\frac{D_{1}^{2}}{2} -D_{1}{\cdot}D_{3}\ .
\end{eqnarray}

\noindent The corresponding Chern characters for the R-sheaves dual the 
Coulomb branes with are:
\begin{eqnarray}
\ch(R^{2}_{(2)})&=& D_{1} +D_{4}-\frac{5D_{1}^{2}}{2} -\frac{3h}{2}\ ,
\nonumber \\
\ch(R^{4}_{(2)})&=& D_{1} +D_{4} +D_{5} +\frac{D_{1}^{2}}{2} -\frac{h}{2}
 -\frac{3f}{2} \ .
\end{eqnarray}

\subsection*{Type II}

The Chern characters of the R-sheaves for the Higgs branes are :
\begin{eqnarray}
\ch(R^{0}_{(2)})&=& D_{3} -\frac{D_{3}^{2}}{2} \ ,\nonumber \\
\ch(R^{1}_{(2)})&=& D_{3} -\frac{3D_{3}^{2}}{2} -D_{3}{\cdot}D_{1}\ .
\end{eqnarray}

\noindent
The Chern characters for the R-sheaves for the Coulomb branes are:
\begin{eqnarray}
\ch(R^{2}_{(2)})&=& D_{3} +D_{4} +D_{5} +\frac{D_{3}^{2}}{2} 
-\frac{h}{2} +\frac{f}{2}\ ,\nonumber \\
\ch(R^{3}_{(2)})&=& D_{3} +D_{5} -\frac{3D_{3}^{2}}{2} 
-h -\frac{7f}{2}\ ,\nonumber \\
\ch(R^{4}_{(2)})&=& D_{3} +D_{5} -h -\frac{9f}{2} -D_{1}{\cdot}D_{3} 
-\frac{3D_{3}^{2}}{2} \ .
\end{eqnarray}
In the above expressions the class of a point has not been shown,
as it is undetermined.  We re-emphasise that in the above
expressions, the terms corresponding to the fibres are fixed only
up to linear equivalences, while the leading terms corresponding
to the divisors are uniquely fixed by the equations.

Now one can try to write down explicit objects which have these
Chern characters. Of course, we don't have the exact expressions
for the complete Chern character. In particular the class of a
point is undetermined, so we cannot hope to retrieve the explicit
objects. However, the R-sheaves corresponding to the Higgs branes
are expected to be given by line bundles with support on the
appropriate non-compact divisor. Looking for line bundles as the
corresponding objects, one can show that one can uniquely write
down such line bundles such that their Chern characters match
with expressions for the R-sheaves corresponding to the Higgs
branes, up to the class of the point. The objects so obtained
have a nice and simple structure. The explicit representations
for the Higgs branch branes are:

\subsection*{ For the $\BC^{3}/\BZ_{3}$ orbifold:} 

\begin{eqnarray}
R^{0}_{(2)}&= i_{*}\left(\mathcal{O}_{D_{1}}\right)\ ,   \\
R^{1}_{(2)}&= i_{*}\left(\mathcal{O}_{D_{1}}(-D_{1})\right)\ , 
\end{eqnarray}
where $i:D_1\rightarrow X$.

\subsection*{The $\BC^{3}/\BZ_{5}$ orbifold} 

\noindent {\bf Type I}
\begin{eqnarray}
R^{0}_{(2)}= i_{*}\left(\mathcal{O}_{D_{1}}\right)\ ,  \\
R^{1}_{(2)}=  i_{*}\left(\mathcal{O}_{D_{1}}(-D_{1})\right)\ , \\
R^{3}_{(2)}= i_{*}\left(\mathcal{O}_{D_{1}}(-D_{3}\right)\ ,
\end{eqnarray} 
where $i:D_1\rightarrow X$.

\noindent {\bf Type II}
\begin{eqnarray}
R^{0}_{(2)}= j_{*}\left(\mathcal{O}_{D_{3}}\right)\ ,  \\
R^{1}_{(2)}= j_{*}\left(\mathcal{O}_{D_{3}}(-D_{1})\right)\ ,
\end{eqnarray}
where $j:D_3\rightarrow X$.
Using certain sequences one can check that the Chern characters of these
objects match with the expressions given above, up to the class of the
point. This has been carried in the appendix C. 

Now for the Coulomb branch branes we have no such guide to write
down the objects, since there is no reason why they should be
line bundles, for instance. Moreover, even if one assumes that
they are line bundles the choice of object is not unique. This is
simply because of the technical fact that there are more terms in
the various expression of the Chern character and there are many
possible ways in which they can be written.

However the crucial point to note is that the structure of the
Chern characters for these Coulomb branch branes is such that
they cannot be written as objects restricted to the appropriate
non-compact divisor.  So they are in general of the form
\begin{equation}
R^{i}_{(2)}= i_{*}A + j_{*}B\ ,
\end{equation}
where $A$ is an object with support on the corresponding non-compact
divisor while $B$ has support on a compact divisor. 
  
In the appendix D, we will nevertheless write down simple representative
objects for these Coulomb branch branes as well. It should be stated
that this is not a unique representation.

\section{Conclusion}

In this paper we have continued further with the study of the quantum
McKay correspondence that was proposed in our previous paper.  To
summarise, we have studied the fractional two-branes in the
$\BC^{3}/\BZ_{3}$ and the $\BC^{3}/\BZ_{5}$ examples in the framework of
toric geometry. 

We have identified the Higgs branch branes as well as the Coulomb
branch branes in these examples. We have discussed further the
quantum McKay correspondence for the fractional 2 branes,
generalising the McKay correspondence for the fractional zero
branes.

We have given the explicit objects for the tautological branes
corresponding to the Higgs branch branes in terms of line
bundles, with support on appropriate non-compact divisors. For
the Coulomb branch branes we see the associated R-sheaves are
objects which cannot be written as objects with support only on
the non-compact divisors.

Our analysis has been based on several choices we made in solving
the related equations for computing the Chern character of the
fractional two-branes as well as the R-sheaves. However we feel
that the very fact that a solution exists with the desired
features is non-trivial. A deeper understanding, both
mathematically and physically, of the quantum McKay
correspondence is desirable.

It would be interesting if a more direct relation of our analysis
could be found to the discussion of Martinec and Moore using some
version of the Hori-Vafa map\cite{HV}.

\acknowledgments
%\section{Acknowledgments}

Bobby Ezhuthachan would like to thank the Institute of
Mathematical Sciences, Chennai, where this work was mainly done,
for support as a Junior Research Fellow. He would also like to
thank the theory groups at HRI, Allahabad, IACS, Kolkata and the
string theory group at TIFR, for the opportunity to present this
work in seminars and useful discussions. S. Govindarajan would
like to thank the Theory Group at CERN for hospitality and
support while some of this work was carried out.  T. Jayaraman
would like to thank the Dept. of Mathematics, TIFR, for
hospitality and support as a Visiting Scientist during the course
of this work and acknowledge very useful discussions with V.
Srinivas.  We would like to thank collectively Ilka Brunner,
Arijit Dey, Subir Mukhopadhyay, Nitin Nitsure, Koushik Ray and
Parameswaran Sankaran, for useful discussions on various
occasions.

\appendix

\section{Some details of the push-forward}

In this appendix, we will explain some of the relevant details of
the push-forward map that has been used in the main text of the
paper. Let us consider the map $i: X \rightarrow Y$. The
push-forward in homology is straightforward -- $i_*: H_k(X)
\rightarrow H_k(Y)$. After all a $k$-cycle remains a $k$-cycle
whether one is on one manifold or the other. What is non-trivial
is the push-forward in cohomology which one can figure out by
using the usual Poincar\'e duality. One first takes the
Poincar\'e dual of a cohomology class in $X$, pushes forward the
homology class by $i_*$ and then takes the Poincar\'e dual again.
If the dimensions of $X$ and $Y$ are $m$ and $n$ respectively,
then $i_* : H^k(X) \rightarrow H^{n-m + k}(Y)$. The next
important fact is to note that
\begin{equation}
\int_X \omega = \int_Y i_* \omega\ .
\end{equation}

Notice that this works only if the entire integrand in the
integral over $Y$ is the $i_*$ of something on $X$. To achieve
this in general we need to use relations that are known in the
mathematical literature as projection formula. (Such projection
formulae are quite important and one needs to use the right one
in context). The one relevant to us is as follows:
\begin{equation}
i_*(E \otimes i^*F) = i_*(E) \otimes F\ ,
\end{equation}
where $E \in H^*(X)$ and $F \in H^*(Y)$. 

By the Grothendieck-Riemann-Roch theorem we have that
\begin{equation}
i_*\big[\ch(E)\ \textrm{Td}(X)\big] = i_*\big[\ch(E)\big]\ \textrm{Td}(Y)\ .
\end{equation}
where $E \in H^*(X)$. Using the GRR theorem and the projection formula,
we can show 
the intersection forms when computed in terms
of local Chern characters in the total space are the same as the ones 
computed directly on the compact divisor.

\section{Toric Geometry - Basics}
\label{toric}

In this appendix, we will briefly review how to construct toric diagrams
for orbifolds as well as to read off various information about the
orbifold space from the toric data. We will discuss the specific
examples of $\BC^{3}/\BZ_{3}$,$\BC^{3}/\BZ_{5}$ orbifolds, which are
discussed in the paper. \subsection{The $\BC^{3}/\BZ_{3}$ orbifold}
First consider the $\BC^{3}/\BZ_{3}$ orbifold with orbifold action
$\frac{1}{3}[1,1,1]$.  In the toric geometry picture this orbifold is
represented by the cone spanned by the vertices
\begin{eqnarray}
v_{1}=(1,0,0),~ v_{2}=(0,1,0),~ v_{3}=(-1,-1,3)\ .
\end{eqnarray}
To see that this cone describes the $\BC^{3}/\BZ_{3}$ orbifold, we first
construct the dual cone. This is done by the following procedure. 
If $(a,b,c)$ is a vector in
the dual cone, then we look for those vectors such that the inner product
of this with each of
the above vertices is positive semidefinite. this gives the following
inequalities.
\begin{eqnarray}
a\ge 0,~ b\ge 0,~ 3c\ge b+a\ .
\end{eqnarray}
Now we have to solve these inequalities to get the basis vectors of the
dual cone. All other solutions to $(a,b,c)$ can be written as a positive
linear combination of these basis vectors and moreover no basis vector
can be expressed as a positive linear combination of any others. For the
example at hand the solutions are given by the following 10 vectors.
\begin{eqnarray}
&&v'_{1}=(0,0,1),~ v'_{2}=(3,0,1),~ v'_{3}=(0,3,1),~ v'_{4}=(2,1,1), \nonumber
\\
&&v'_{5}=(1,2,1),~ v'_{6}=(1,1,1),~ v'_{7}=(2,0,1),~ v'_{8}=(0,2,1), \nonumber
\\
&&v'_{9}=(1,0,1),~ v'_{10}=(0,1,1)\ .
\end{eqnarray}
Each of these vectors is associated with a monomial. For
example
\begin{eqnarray}
v'_{1}\equiv Z,~ v'_{2}\equiv X^{3} Z\ .
\end{eqnarray}
Now we will digress a bit. Consider polynomials in two variables
$(U,V)$. Then the domain over which these arbitrary polynomials are well
defined ,which we denote by $\BC[U,V]$, is actually $\BC^{2}$, so
$\BC[U,V]$ is the coordinate ring of $\BC^{2}$.  We will use the
shorthand notation $\BC[U,V]\equiv \BC^{2}$.  Similarly if we look at
the domain over which polynomials of the variables $(U,V,U^{-1},V^{-1})$
are well defined it describes the space $(\BC^{*})^{2}$, because the
functions are not defined at $(U,V)=(0,0)$. Similarly if we consider
polynomials in three variables $(U,V,W)$, then $\BC[U,V,W]\equiv
\BC^{3}$. The orbifold $\BC^{3}/\BZ_{3}$ with orbifold action
$\frac{1}{3}$[1,1,1] on $(U,V,W)$ can be described as the domain over
which all polynomials constructed out of variables, which are single
valued on the orbifold, is defined. Therefore,
\begin{equation}
\BC^{3}/\BZ_{3}\equiv
\BC[U^{3},V^{3},W^{3},UVW,UV^{2},VU^{2},VW^{2},WV^{2},UW^{2},WU^{2}]\ .
\end{equation}

Now we can see how to read off the space from the data we obtained from
the dual cone. Writing the monomial associated to each of the dual basis
vectors we construct the domain over which polynomials with these
monomials as the variables are well defined. This in our notation is
written as 
$$
\BC[Z,X^{3}Z,Y^{3}Z,X^{2}YZ,XY^{2}Z,XYZ,X^{2}Z,Y^{2}Z,XZ,YZ]\ .
$$
After
changing variables to $X=\frac{U}{W}$, $Y=\frac{V}{W}$\ \textrm{and}\
$Z=W^{3}$, we get 
$$
\BC[W^{3},U^{3},V^{3},U^{2}V,V^{2}U,UVW,U^{2}W,V^{2}W,UW^{2},VW^{2}]
\ .
$$
This is the description of $\BC^{3}/\BZ_{3}$, that we saw earlier.

\subsubsection{Resolution of the orbifold}

To resolve the orbifold, the strategy is to subdivide the cone into
several smaller cones by inserting more vectors in the interior of the
cone such that for each sub-cone the determinant of the generators of
that particular cone, which is also the volume of the particular cone,
is one. One can easily see that this criteria is not satisfied by the
original cone itself. For the the $\BC^{3}/\BZ_{3}$ orbifold, this is
achieved by taking one more vector
\begin{equation}
v_{4}=(0,0,1)\ ,
\end{equation}
which subdivides the cone to three sub-cones, each of which have a unit
determinant. The new cone so obtained is given in Figure \ref{fig1} Now
as before construct the dual cones for each of the cones, and as before
we have the following inequalities.
\begin{center}
\begin{tabular}{|c|c|c|}\hline
Cone 1 & Cone 2  & Cone 3 \\ \hline
$c\ge 0, ~a\ge 0, ~3c\ge a+b$ & $c\ge 0, ~b\ge 0, 
~3c\ge a+b$ &  $c\ge 0, ~a\ge 0, ~b\ge 0$ \\
$v'_{1}=(0,3,1)$, & $v'_{1}=(3,0,1)$,
& $v'_{1}=(1,0,0)$, \\
$ v'_{2}=(0,-1,0)$, & $ v'_{2}=(-1,0,0),$ & $ v'_{2}=(0,1,0)$,  \\
$v'_{3}=(1,-1,0)$ & $v'_{3}=(-1,1,0)$ & $v'_{3}=(0,0,1)$ \\
$\BC[Y^{3}Z, Y^{-1}, XY^{-1}]$ & $\BC[X^{3}Z, X^{-1}, 
YX^{-1}]$ & $\BC[X, Y, Z]$ \\
\hline
\end{tabular}
\end{center}
The divisor corresponding to $v_{4}$ is given by $Z=0$ and is
obtained by substituting $Z=0$ in the above. Then one has the
following spaces $\BC[X,Y]$, $\BC[X^{-1},YX^{-1}]$,
$\BC[Y^{-1},XY^{-1}]$. These are to be thought of as local
coordinate patches of some space. What space do these patches
describe? They describe the space $\BP^{2}$. This can be seen by
looking at the patches of $\BP^{2}$. $\BP^{2}$ is given by
$(U,V,W)\sim (\lambda U,\lambda V,\lambda W)$. Then we have three
patches given by the regions where $U, V, W$ are individually non
zero. In each of these patches the coordinates can be taken to be
$(\frac{V}{U},\frac{W}{U})$, $(\frac{U}{V}, \frac{W}{V})$,
$(\frac{U}{W}, \frac{V}{W})$. Defining $X=\frac{U}{W}$ and
$Y=\frac{V}{W}$, we have the following three patches $(X,Y)$,
$(X^{-1},YX^{-1})$, $(Y^{-1},XY^{-1}),$ so comparing with what we
got from the toric analysis we see that the space after
resolution is indeed a $\BP^{2}$ so we see that ${\bf D_{4}\equiv
\BP^{2}}$.

\subsection{The $\BC^{3}/\BZ_{5}$ orbifold}

Now we consider the example of the $\BC^{3}/\BZ_{5}$ orbifold with
orbifold action $\frac{1}{5}$[1,1,3]. The vertices for the cone are
given by
\begin{equation}
v_{1}=(1,0,0),\ v_{2}=(-1-3,5),\ v_{3}=(0,1,0) \ .
\end{equation}
Using the same method outlined before one can
check that this is indeed the $\BC^{3}/\BZ_{5}$ orbifold.

\subsubsection{The resolution of $\BC^{3}/\BZ_{5}$}

Following the process for resolution as described in the
$\BC^{3}/\BZ_{3}$ example one finds that one has to insert two vertices
\begin{eqnarray}
v_{4}=(0,-1,2),\ v_{5}=(0,0,1), ~~~~~\textrm{see Figure \ref{fig2}}
\end{eqnarray}
inside the cone to get the desired condition of unit determinant for the
individual sub-cones. Now as before construct the dual fans for each of
the cones, and as before we have the following inequalities.
\begin{eqnarray}
\textrm{Cone 1:}\ c\ge 0,\ a\ge 0\ \textrm{and}\ b\ge 0 \ .
\end{eqnarray}
The corresponding vertices are given by
\begin{eqnarray}
v'_{1}=(1,0,0), ~v'_{2}=(0,1,0), ~v'_{3}=(0,0,1)\ .
\end{eqnarray}
The space is given by $\BC[X,Y,Z]$
\begin{eqnarray}
\textrm{Cone 2:}\ c\ge 0,\ 5c\ge 3b+a\ \textrm{and}\ b\ge 0\ .
\end{eqnarray}
The corresponding vertices are given by
\begin{eqnarray}
v'_{1}=(-1,0,0),~ v'_{2}=(-3,1,0),~ v'_{3}=(5,0,1)\ .
\end{eqnarray}
The space is given by $\BC[X^{-1},X^{-3}Y,X^{5}Z]$.
\begin{eqnarray}
\textrm{Cone 3:}\  c\ge 0,\ 2c\ge b\ \textrm{and}\ 5c\ge 3b+a\ .
\end{eqnarray}
The corresponding vertices are given by
\begin{eqnarray}
v'_{1}=(3,-1,0), v'_{2}=(-1,2,1), v'_{3}=(-1,0,0)\ .
\end{eqnarray}
The space is given by $\BC[X^{3}Y^{-1},Y^{2}ZX^{-1},X^{-1}]$.
\begin{eqnarray}
\textrm{Cone 4:}\ 2c\ge b,\ a\ge 0\ \textrm{and}\ 5c\ge 3b+a\ .
\end{eqnarray}
The corresponding vertices are given by
\begin{eqnarray}
v'_{1}=(0,5,3),~ v'_{2}=(1,-2,-1),~ v'_{3}=(0,-2,-1)\ .
\end{eqnarray}
The space is given by $\BC[Y^{5}Z^{3},XY^{-2}Z^{-1},Y^{-2}Z^{-1}]$.
\begin{eqnarray}
\textrm{Cone 5:}\  c\ge 0, 2c\ge b\ \textrm{and}\ a\ge 0\ .
\end{eqnarray}
The corresponding vertices are given by
\begin{eqnarray}
v'_{1}=(0,-1,0),~ v'_{2}=(1,0,0),~ v'_{3}=(0,2,1)\ .
\end{eqnarray}
The space is given by $\BC[Y^{-1},X,Y^{2}Z]$.

Now the divisor $\textrm{D}_{4}$ corresponding to $v_{4}$ is given by
$\textrm{D}_{4}\equiv Z^{2}/Y=0$. To find out what space this divisor
corresponds to one has to analyse all the cones of which this is a
common point. These will be the coordinate patches of the corresponding
space. Since there are three cones surrounding this point, the
corresponding space should be a $\BP^{2}$. This can be checked
rigorously, exactly as before. To do this we substitute
$\textrm{D}_{4}=0$ in the cones (3), (4) and (5). Take $Y^{2}Z\equiv A$
and $\textrm{D}_{4}=0$. We then get for the corresponding patches,
$\BC[AX^{-1}, X^{-1}]$, $\BC[A,X]$ and $\BC[XA^{-1},A^{-1}]$
respectively. As noted earlier these are the patches of $\BP^{2}$, so
\begin{eqnarray}
\textrm{D}_{4}\equiv \BP^{2}\ .
\end{eqnarray}
To find the space corresponding to $\textrm{D}_{5}$ given by $Z=0$ we
similarly substitute $Z=0$ in the patches (1),(2),(3),(5). We then get
for the corresponding patches:
\begin{eqnarray}
(1)~ \BC[X,Y],~ (2)~ \BC[X^{-1}, X^{-3}Y],~ (3) ~\BC[X^{3}Y^{-1}, X^{-1}],~ (5)~ \BC[Y^{-1},X]
\end{eqnarray}
These are the coordinate patches associated with the space $\BF_{3}$.
In general the space $\BF_{a}$ has the following four coordinate 
patches\cite{Fulton}.
\begin{eqnarray}
(X^{-1},X^{a}Y), (X,Y), (X^{-1},X^{-a}Y^{-1}), (X,Y^{-1}) \nonumber
\end{eqnarray}
So we see that
\begin{equation}
\textrm{D}_{5}\equiv \BF_{3}\ .
\end{equation}

The general rule for computing the triple intersections is that the
triple intersections for any three distinct divisors which span a cone
is 1 and the intersection of those distinct divisors that don't span a
cone, vanish.  For instance for the $\BC^{3}/\BZ_{5}$ orbifold, the
vectors $v_{1}, v_{2}, v_{4}$ span the cone labelled $4$. So we have
$D_{1}\cdot D_{4}\cdot D_{2}=1$.  On the other hand since $v_{1},
v_{2}, v_{5}$ do not span a cone, we have
$D_{1}\cdot D_{5}\cdot D_{2}=0$, and so on. These triple
intersections, involving the distinct divisors, can then be used to
obtain other triple intersections, which involve self intersections of
divisors. This is done by using the linear equivalence relations between
the divisors to express these in terms of the ones involving distinct
ones.

\section{The R-sheaves for the Higgs branes}
\label{seq1}

We will provide some details that go into computing the Chern character for
the R-sheaves\footnote{Here we are treating both compact and non-compact divisors on par and we will do so throughout this section. A detailed technical justification of this is beyond the scope of this paper.}.  The general sequence we will be using is of the form,
\begin{equation}
0 \rightarrow \mathcal {O}_{X}(-D_{i}-D_{j})
 \rightarrow \mathcal{O}_{X}(-D_{j}) \rightarrow
i_{*}\left(\mathcal{O}_{D_{i}}(-D_{j})\right)\rightarrow 0
\end{equation}
where $i:D_i\rightarrow X$. From the above sequence we obtain
\begin{equation}
\ch\left[i_{*}\left(\mathcal{O}_{D_{i}}(-D_{j})\right)\right] = 
\ch\left[\mathcal{O}_{X}(-D_{j})\right] -\ch\left[\mathcal {O}_{X}(-D_{i}-D_{j}
)\right]
\end{equation}

\subsection*{The R-sheaves for the Higgs branes of the $\BC_{3}/\BZ_{3}$ orbifold}

\begin{equation}
0 \rightarrow \mathcal {O}_{X}(-D_{1}) \rightarrow 
\mathcal{O}_{X} \rightarrow i_{*} 
\left( \mathcal{O}_{D_{1}} \right)
\rightarrow 0
\end{equation}
Using the above sequence can show that, 
\begin{equation}
\ch\left[i_{*} \left(\mathcal{O}_{D_{1}}\right)\right] = \ch\left[R^{0}_{2}\right],  
\end{equation}
up to the class of a point, which was undetermined in the main text.
Similarly for the other R-sheaf, consider the following sequence
\begin{equation}
0 \rightarrow \mathcal {O}_{X}(-2D_{1}) \rightarrow 
\mathcal{O}_{X}(-D_{1}) \rightarrow i_{*}\left(
\mathcal{O}_{D_{1}}(-D_{1})\right)
\rightarrow 0
\end{equation}
From this sequence one can compute the Chern character of 
$\mathcal{O}_{D_{1}}(-D_{1})$ and again up to the 
class of a point, 
\begin{equation}
\ch\left[i_{*} \left(\mathcal{O}_{D_{1}}(-D_{1})\right)\right] = \ch\left[R^{1}_{2}\right],
\end{equation}

\subsection*{The R-sheaves for the Higgs  branes of the $\BC^{3}/\BZ_{5}$
orbifold}

One can carry out a similar exercise for the $\BC^{3}/\BZ_{5}$ orbifold. We
will give below the sequences used to show that the Chern characters of
the objects match with the Chern characters of the R-sheaves. 

\noindent {\bf Type I}

The sequences of interest are,
\begin{eqnarray}
&& 0 \rightarrow \mathcal {O}_{X}(-D_{1}) 
\rightarrow \mathcal{O}_{X} \rightarrow 
i_{*}\left(\mathcal{O}_{D_{1}}\right)\rightarrow 0 \nonumber \\
&&  0 \rightarrow \mathcal {O}_{X}(-2D_{1}) 
\rightarrow \mathcal{O}_{X}(-D_{1}) \rightarrow
i_{*}\left(\mathcal{O}_{D_{1}}(-D_{1})\right)\rightarrow 0 \nonumber \\
&&  0 \rightarrow \mathcal {O}_{X}(-D_{1} -D_{3})
 \rightarrow \mathcal{O}_{X}(-D_{3}) \rightarrow
i_{*}\left(\mathcal{O}_{D_{1}}(-D_{3})\right)\rightarrow 0 
\end{eqnarray}
Using these sequences one can show that up to the class of a point,
\begin{eqnarray}
\ch\left[i_{*}\left(\mathcal{O}_{D_{1}}\right)\right]&=& 
\ch\left[R^{(0)}_{2}\right]\ ,   \nonumber \\
\ch\left[i_{*}\left(\mathcal{O}_{D_{1}}(-D_{1})\right)\right] 
&=& \ch\left[R^{(1)}_{2}\right]\ , \nonumber \\
\ch\left[i_{*}\left(\mathcal{O}_{D_{1}}(-D_{3})\right)\right]
&=& \ch\left[R^{(3)}_{2}\right]\ .
\end{eqnarray}

\noindent {\bf Type II}

\noindent
The sequences of interest are,
\begin{eqnarray}
&& 0 \rightarrow \mathcal {O}_{X}(-D_{3}) \rightarrow \mathcal{O}_{X} 
\rightarrow
j_{*}\left(\mathcal{O}_{D_{3}}\right)\rightarrow 0 \nonumber \\
&&  0 \rightarrow \mathcal {O}_{X}(-D_{1} -D_{3}) \rightarrow 
\mathcal{O}_{X}(-D_{1}) \rightarrow
j_{*}\left(\mathcal{O}_{D_{3}}(-D_{1}\right)\rightarrow 0 
\end{eqnarray}
From the above sequences, one gets,
\begin{eqnarray}
\ch\left[j_{*}\left(\mathcal{O}_{D_{3}}\right)\right] 
&=& \ch\left[R^{(0)}_{2}\right] \ ,  \nonumber \\
\ch\left[j_{*}\left(\mathcal{O}_{D_{3}}(-D_{1}\right)\right] 
&=& \ch\left[R^{(1)}_{2}\right] \ .
\nonumber 
\end{eqnarray}

\section{The Coulomb branes}

In this appendix we will write down some representative objects
for the Coulomb branes. These can be derived in the same way as
the duals for the Higgs branes using appropriate sequences, which
we don't write down. We write them down as sum of terms each
corresponding to objects with support on both the non-compact
divisor and on some compact divisor. We emphasise that these are
not unique. The key point is that unlike the duals for the Higgs
branes, these cannot be written as purely line bundles supported
on the non-compact divisor.

\subsection*{\bf Coulomb brane for the $\BC^{3}/\BZ_{3}$ orbifold}
\begin{equation}
R^{3}_{(2)} = i_{*}\left[\mathcal{O}_{D_{1}}(D_1)\right]
+ j_{*}\left[\mathcal{O}_{D_{4}}(-2D_1)\right]\ .
\end{equation} 

\subsection*{\bf Coulomb branes for the $\BC^{3}/\BZ_{5}$ orbifold}

\noindent {\bf Type I}
\begin{eqnarray}
R^{2}_{(2)} &=& i_{*}\left[\mathcal{O}_{D_{1}}(-2D_{1})\right]
+ j_{*}\left[\mathcal{O}_{D_{4}}(D_4)\right]\ , \\
R^{4}_{(2)} &=& 
i_{*}\left[\mathcal{O}_{D_{1}}(D_1)\right]+j_{*}\left[\mathcal{O}_{D_{4}}(-D_1)\right] 
 +k_{*}\left[\mathcal{O}_{D_{5}}(D_1+D_5)\right]\ . 
\end{eqnarray}

\noindent
{\bf Type II}
\begin{eqnarray}
R^{2}_{(2)} &=& l_{*}\left[\mathcal{O}_{D_{3}}(D_3)\right] 
+ j_{*}\left[\mathcal{O}_{D_{4}}(D_4)\right]+k_{*}\left[\mathcal{O}_{D_{5}}(-2D_1)\right]\ , \\
R^{3}_{(2)} &=&
l_{*}\left[\mathcal{O}_{D_{3}}(-D_{3})\right] 
+ k_{*}\left[\mathcal{O}_{D_{5}}(D_5-D_1)\right] 
\ , \nonumber \\
R^{4}_{(2)}&=& l_{*}\left[\mathcal{O}_{D_{3}}(-D_{3}-D_1)\right] 
+ k_{*}\left[\mathcal{O}_{D_{5}}(D_5-2D_1)\right] \ .
\end{eqnarray}
Here $i_{*}$,  $j_{*}$,  $k_{*}$ and  $l_{*}$ are the push-forwards for the 
appropriate divisors.

\end{document}